\documentclass[prl,twocolumn,floatix,amssymb,showpacs,amsmath,superscriptaddress]{revtex4-1}
\pdfoutput=1
\usepackage{graphicx}
\usepackage{epsfig}
\usepackage{amsmath,amssymb}
\usepackage{ dsfont }
\usepackage{color}

\DeclareMathOperator{\Tr}{Tr}

\begin{document}

\newcommand{\ket}[1]{\ensuremath{\left|{#1}\right\rangle}}
\newcommand{\bra}[1]{\ensuremath{\left\langle{#1}\right|}}
\newcommand{\quadr}[1]{\ensuremath{{\not}{#1}}}
\newcommand{\quadrd}[0]{\ensuremath{{\not}{\partial}}}
\newcommand{\slpar}{\partial\!\!\!/}
\newcommand{\gtrescero}{\gamma_{(3)}^0}
\newcommand{\gtresuno}{\gamma_{(3)}^1}
\newcommand{\gtresi}{\gamma_{(3)}^i}

\title{Many-Body Interactions with Tunable-Coupling Transmon Qubits}

\date{\today}

\author{A. Mezzacapo}
\affiliation{Department of Physical Chemistry, University of the Basque Country
UPV/EHU, Apartado 644, E-48080 Bilbao, Spain}
\author{L. Lamata}
\affiliation{Department of Physical Chemistry, University of the Basque Country
UPV/EHU, Apartado 644, E-48080 Bilbao, Spain}
\author{S. Filipp}
\affiliation{Department of Physics, ETH Z\"urich, CH-8093 Z\"urich, Switzerland}
\author{E. Solano} 
\affiliation{Department of Physical Chemistry, University of the Basque Country
UPV/EHU, Apartado 644, E-48080 Bilbao, Spain}
\affiliation{IKERBASQUE, Basque Foundation for Science, Alameda Urquijo 36, 48011
Bilbao, Spain}

\begin{abstract}
The efficient implementation of many-body interactions in superconducting circuits allows for the realization of multipartite entanglement and topological codes, as well as the efficient simulation of highly correlated fermionic systems. We propose the engineering of fast multiqubit interactions with tunable transmon-resonator couplings. This dynamics is obtained by the modulation of magnetic fluxes threading superconducting quantum interference device loops embedded in the transmon devices. We consider the feasibility of the proposed implementation in a realistic scenario and discuss potential applications.\end{abstract}


\maketitle
Superconducting qubits coupled to transmission line resonators have proved to be physical systems well suited for quantum information processing~\cite{Wallraff04,Devoret13}. The coherent control performed on this kind of device at the quantum level has produced a series of remarkable results~\cite{Steffen06, Fedorov12,Steffen13}. It has been proven that this quantum platform can reach ultrastrong-coupling regimes~\cite{Niemczyk10,Peropadre10}. Among superconducting qubits, transmon qubits are currently the most robust and reliable. They are designed in order to suppress offset charge noise to negligible values~\cite{Koch07}. Protocols of quantum information have been implemented, such as error correction up to three qubits~\cite{Reed12} and experimental tests of fundamental quantum mechanics~\cite{Abdumalikov13}. Implementations of quantum simulators of spin and coupled spin-boson systems have been recently proposed~\cite{Heras14,Mezzacapo14}. Complex entangled states encoded in superconducting transmon qubits have already been proposed and realized experimentally~\cite{Bishop09,Neeley10,DiCarlo10}. However, state-of-the-art realizations of many-qubit entangled states still rely on complex sequences of gates, and implementations of effective many-body interactions represent a tough challenge.

The introduction of collective entangling operations in superconducting devices can ease several tasks of quantum information processing. They have been proposed theoretically~\cite{Molmer} and realized experimentally in ion traps up to fourteen qubits~\cite{Monz11}. Similarities between ion-trap systems and superconducting circuits have been already investigated~\cite{Liu07}. By means of collective gates, one can drive the generic many-qubit transition $|00\cdots0\rangle\rightarrow|11\cdots1\rangle$ and prepare multipartite Greenberger-Horne-Zeilinger states with a single operation. The transition can be obtained with effective simultaneous red and blue sidebands acting upon the ions. The latter have been also demonstrated in a variety of superconducting setups~\cite{Wallraff07, Leek08, Strand13}. Sequences of  collective gates, together with local qubit rotations, can i andmplement stabilizer operators~\cite{Mueller11,Nigg12}, that can allow for the implementation of topological codes~\cite{Kitaev03}. Recently it has been shown that collective qubit interactions allow for efficient simulation of fermionic dynamics and coupled fermionic-bosonic systems~\cite{Casanova12,Mezzacapo12}.

In this Letter, we propose the implementation of effective many-body interactions among several tunable-coupling transmons inside a microwave cavity. We consider three-island superconducting devices~\cite{Gambetta11,Srinivasan12}, addressed as tunable-coupling transmon qubits (TCQs), coupled to a coplanar microwave resonator. Then, we show that dynamically sweeping flux biases, acting on two SQUID loops embedded in the three-island devices, it is possible to perform simultaneous red and blue-sideband transitions of many qubits. This leads to effective collective entangling gates that can be used to efficiently obtain many-particle operators. We demonstrate that the third level of the single TCQ can be ruled out of the dynamics. Finally, we validate the proposal with numerical simulations of the system dynamics taking into account a realistic decoherence model.

\begin{figure}
\includegraphics[scale=0.34]{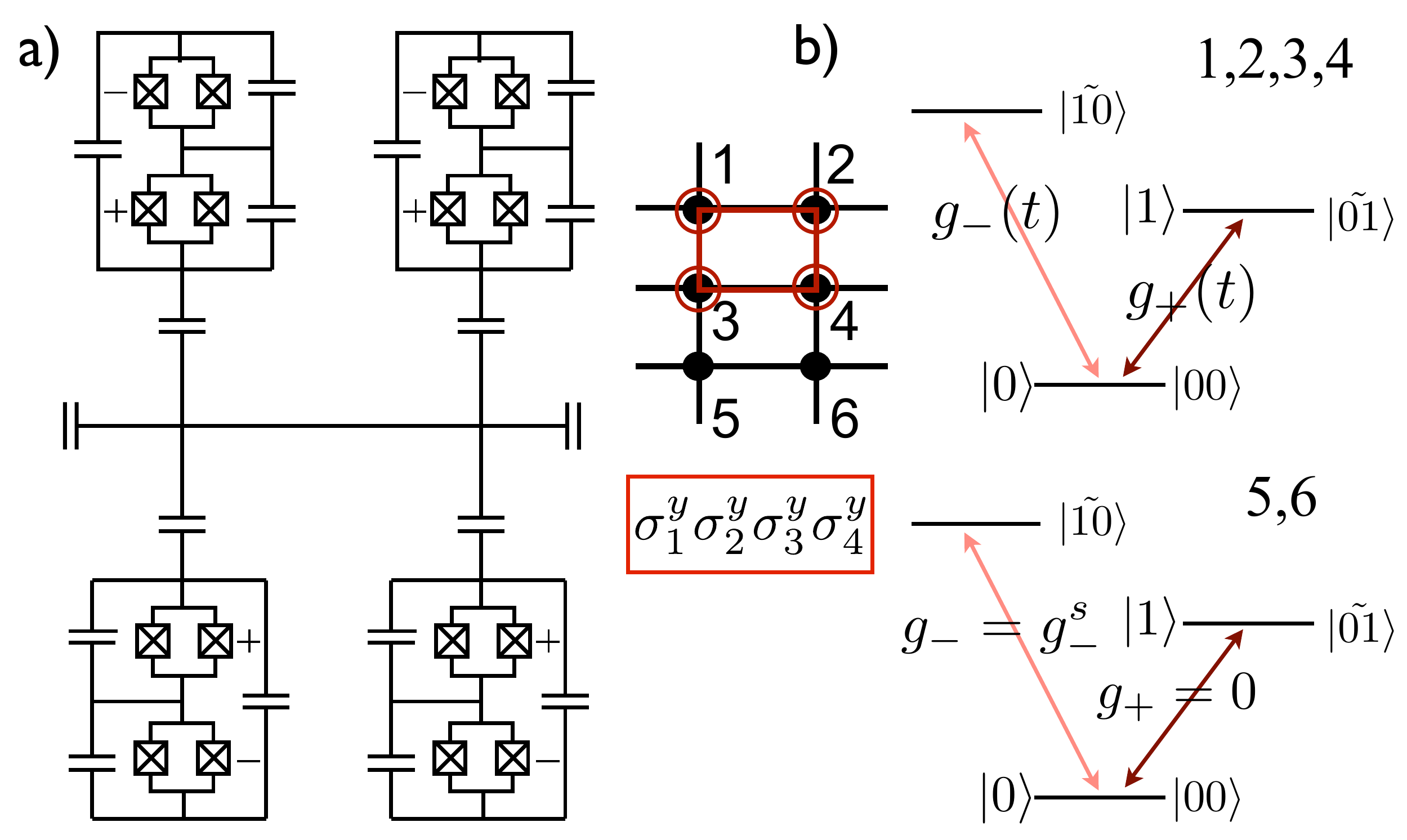}
\caption{(Color online) a) Scheme of a setup composed of four TCQs capacitively coupled to a coplanar resonator. The SQUID loops labeled with $+$ and $-$ can be threaded by external magnetic fluxes. b) Generation of many-particle operator $\sigma_1^y \sigma_2^y \sigma_3^y\sigma_4^y$, between the first, second, third and fourth qubits~\cite{Suppl}. Selectivity is obtained by setting the coupling of the other qubits to the resonator to zero. The qubit logical levels $|0\rangle$ and $|1\rangle$ are the first levels of the TCQ, $|00\rangle$ and $|\tilde{01}\rangle$.    \label{circuitscheme}} 
\end{figure}

We start by considering a setup made of a resonator coupled to several TCQs, as in Fig.~\ref{circuitscheme}a. We show that under specific conditions, the TCQs in the setup behave as two level systems and the effective interaction among them is given by the Hamiltonian
\begin{equation}
H_{I_{\textrm{eff}}}=-\xi\sum_{i<j} \sigma_i^\alpha\sigma_j^\alpha.\label{HamMS}
\end{equation}
Here, $\xi$ is the interaction strength that sets the speed of the transition and the Pauli matrix $\sigma_i^\alpha$, with either $\alpha=x$ or $\alpha=y$, refers to the subspace spanned by the two lowest energy levels of the $i$-th TCQ. A single device is composed of three superconducting islands: the upper and lower islands are connected to a central one by means of two SQUID loops. Their effective Josephson couplings $E_{J_\pm}(\Phi_\pm)$ can be tuned by threading the respective superconducting loops with external magnetic fluxes $\Phi_\pm$.  In the symmetric limit for the two Josephson junctions of the loops, one has $E_{J_\pm}=E_{J_\pm}^M\cos(\pi\Phi_\pm/\Phi_0)$, where $E_{J_\pm}^M$ is the total Josephson energy of the junctions, $\Phi_0$ being the fundamental flux quantum. The Hamiltonian of the individual TCQ, neglecting the interaction with the resonator, reads $H_T=\sum_\pm 4E_{C_\pm}(n_{\pm}-n'_{g_{\pm}})^2-\sum_{\pm}E_{J_{\pm}}\cos(\gamma_\pm)+4E_In_+n_-$.  Here, $\gamma_\pm$ are the gauge invariant phase differences on the upper and lower SQUID loops, $n_\pm$ the charge associated, with the offset charge due to gate voltage bias $n'_{g_{\pm}}$. The charging energies of the upper and lower islands are labeled by  $E_{C_\pm}$, while $E_I$ stands for the interaction energy between them. In the limit $E_{J_\pm}\gg E_{C_\pm}$, the charge dispersion of the device is negligible~\cite{Koch07}. One can expand to fourth order the cosine potentials associated with the Josephson energies and write the Hamiltonian as a coupled anharmonic oscillator model, $H_{T_{\textrm{eff}}}=\sum_\pm [\omega_\pm+\delta_\pm(b_\pm^{\dagger}b_\pm-1)/2]b^{\dagger}_\pm b_\pm+J(b_+b^{\dagger}_-+b_+^{\dagger}b_-)$. Here, and in the following, we have set $\hbar=1$. The anharmonicity factors depend on the charging energies $\delta_\pm=-E_{C_\pm}$ and the parameters $\omega_\pm$, $\delta_\pm$ and $J$ are defined in terms of the two external flux biases $\Phi_\pm$~\cite{Gambetta11}. 

We consider here that the two external fluxes are changed in time, with some time-dependent functions $\Phi_\pm(t)$. While the fluxes change in time, the parameters in the Hamiltonian $H_{T_{\textrm{eff}}}$ follow accordingly. We apply to $H_{T_{\textrm{eff}}}$ the time-dependent unitary  $T(t)=e^{\lambda(t)(b_+b_-^{\dagger}-b_+^{\dagger}b_-)}$, where the phase $\lambda(t)$ is defined instantaneously as a function of the parameters of the time-dependent Hamiltonian $H_{T_{\textrm{eff}}}$. The resulting transformed Hamiltonian $\tilde{H}_{T_d}=T^{\dagger}(t)H_{T_{\textrm{eff}}}T(t)-iT^{\dagger}(t)\dot{T}(t)$ reads 
\begin{align}
\label{HTd}
\tilde{H}_{T_d}=&\sum_{\pm}\left[\tilde{\omega}_{\pm}+\frac{\tilde{\delta}_{\pm}}{2}(\tilde{b}_\pm^{\dagger}\tilde{b}_\pm-1)\right]\tilde{b}_\pm^{\dagger}\tilde{b}_\pm+\nonumber \\
&\tilde{\delta}_c\tilde{b}_+^{\dagger}\tilde{b}_+\tilde{b}_-^{\dagger}\tilde{b}_- 
+i\dot{\lambda}(t)(\tilde{b}^{\dagger}_+\tilde{b}_- -\tilde{b}_+ \tilde{b}^{\dagger}_-).
\end{align}
One can recognize in the above Hamiltonian a diagonal part and an off-diagonal term that results in a small renormalization of the energy levels. The diagonal part reads
\begin{equation}
\tilde{H}_0=\sum_{\pm}\left[\tilde{\omega}_{\pm}+\tilde{\delta}_{\pm}/2(\tilde{b}_\pm^{\dagger}\tilde{b}_\pm-1)\right]\tilde{b}_\pm^{\dagger}\tilde{b}_\pm+
\tilde{\delta}_c\tilde{b}_+^{\dagger}\tilde{b}_+\tilde{b}_-^{\dagger}\tilde{b}_-.
\end{equation}
The first two excited levels of $\tilde{H}_0$ are defined by the occupation of the two modes $\tilde{b}^{\dagger}_\pm$ and have energies $\tilde{\omega}_\pm$.
When the two external magnetic fluxes $\{\Phi_+(t),\Phi_-(t)\}$ are driven in time, the first two excited levels of the Hamiltonian $\tilde{H}_0$ are continuously sweeping between different states in the original basis, as $|\tilde{01}\rangle=\tilde{b}_+^{\dagger}|00\rangle=\cos(\lambda)|01\rangle+\sin(\lambda)|10\rangle$,  $|\tilde{10}\rangle=\tilde{b}_-^{\dagger}|00\rangle=\cos(\lambda)|01\rangle-\sin(\lambda)|10\rangle$. One can use the two levels $|0\rangle \equiv |00\rangle$ and $|1\rangle \equiv |\tilde{01}\rangle$ as a qubit, see Fig.~\ref{circuitscheme}b.

We focus now on the interaction term between a single TCQ and the resonator, when the flux biases are varied in time. The TCQs are capacitively coupled to a coplanar resonator, of frequency $\omega_r$. Their interaction can be modeled as $H_I=2eV_{\textrm{\textrm{rms}}}(\beta_+n_+ +\beta_- n_-)(-ia^{\dagger}+ia)$, where the $a$, $a^{\dagger}$ operators act on the resonator field.  The coupling prefactors $\beta_\pm$ are defined by the circuit capacitances, while $V_{\textrm{\textrm{rms}}}$ stands for the root mean square voltage of the resonator. We consider identical capacitances for the upper and lower islands ($\beta_\pm=\beta$). Non-symmetric capacitance configurations do not change the nature of the problem and result in small deviations in the numerical analysis~\cite{Gambetta11}. The interaction can be expressed in the frame of $T(t)$, 
\begin{equation}
\tilde{H}_I=\sum_\pm g_\pm(t)(\tilde{b}^{\dagger}_\pm -\tilde{b}_\pm)(a^{\dagger}-a).
\end{equation}
We introduce a two-tone driving of the coupling $g_+(t)\equiv 2eV_{\textrm{rms}}\beta\langle1|\hat{n}|0\rangle$, with $\hat{n}=n_++n_-$, between the first two levels of the TCQ and prove later that it can be realized by proper flux drivings,   
\begin{equation}
g_+(t)\equiv g_{+}^s+g_+^d[\cos(\omega_g t)+\cos(\omega'_g t)].\label{tunablecoupling}
\end{equation}  
Here, we have defined a static contribution $g_{+}^s$ and a dynamical part, where $g_+^d$ sets the strength of the two-tone $\omega_g,\omega'_g$ modulation.  The frequencies of the coupling are chosen to be detuned by $\delta$ with respect to the qubit-resonator sidebands,  $\omega_g=\omega_r +\tilde{\omega}_+-\delta$ and $\omega'_g=\omega_r -\tilde{\omega}_+ -\delta$.
Namely, in interaction picture with $\tilde{H}_0$, the effective TCQ-resonator Hamiltonian can be written as~\cite{Suppl}
\begin{equation}
\tilde{H}_I=  \tilde{H}_{I_{\textrm{JC}}} + \tilde{H}_{I_+} + \tilde{H}_{I_-} .\label{3Ham}
\end{equation}
The first term of this Hamiltonian is a Jaynes-Cummings interaction due to the static contributions to the couplings $g_\pm(t)$, $\tilde{H}_{I_{\textrm{JC}}}=-\sum_\pm g_\pm^s(\tilde{b}_\pm a^{\dagger}+\tilde{b}_\pm^{\dagger} a)$, which results in an effective interaction of coupling strength $(g^s_\pm)^2/\Delta_\pm$, where $\Delta_\pm=\tilde{\omega}_\pm-\omega_r$ is the detuning of the first two TCQ levels from the resonator frequency.
The second and third terms of the right side of Eq.~(\ref{3Ham}) $\tilde{H}_{I_\pm}$ involve the dynamical contribution to the coupling terms, proportional to $g_\pm^d$. The term acting on the first two levels, imposing the condition of Eq.~(\ref{tunablecoupling}), reads
\begin{equation}
\tilde{H}_{I_+}=g_+^d\big[\cos(\omega_g t) + \cos(\omega'_g t)\big](\tilde{b}^{\dagger}_+ -\tilde{b}_+)(a^{\dagger}-a).\label{qubitHam}
\end{equation}
Neglecting fast oscillating terms, Eq.~(\ref{qubitHam}) reduces to $\tilde{H}_{I_+}\approx ig_+^d/2\bigg(a^{\dagger}e^{i\delta t}-ae^{-i\delta t}\bigg)\sigma^y$, where $\sigma^y$ is a Pauli matrix acting on the Hilbert space spanned first the two levels of the device. The third contribution to the dynamics, $\tilde{H}_{I_-}$ in Eq.~(\ref{3Ham}), has several terms oscillating at different frequencies. If none of them is close to the third level sidebands, contribution from $\tilde{H}_{I_-}$ will be negligible and leakage to the third level will be suppressed. 
In fact, when the dynamical detuning is much smaller than qubit-resonator one, $(g_{\pm}^s)^2/\Delta_\pm \ll (g^d_+)^2/4\delta$, the dynamics will be dominated by $\tilde{H}_{I_+}$. A small Stark-Lamb shift term $\sum_{j} (g_+^s)^2/\Delta_+\sigma_j^z(\frac{1}{2}+a^{\dagger}a)$, can be considered negligible, taking into account small cavity population and renormalization of the qubit frequencies. 
Provided with TCQ-resonator interactions as in Eq.~(\ref{3Ham}), one can build multi-qubit setups, where the effective total Hamiltonian reads
\begin{equation}
\tilde{H}_{I_\textrm{eff}}=\sum_j i\frac{g_b}{2}\bigg(a^{\dagger}e^{i\delta t}-ae^{-i\delta t}\bigg)\sigma_j^y,\label{Hresonant} and
\end{equation}
where $\sigma_j^y$ refers to the first two levels of the $j$-th TCQ.
\begin{figure}\centering
\includegraphics[scale=0.16]{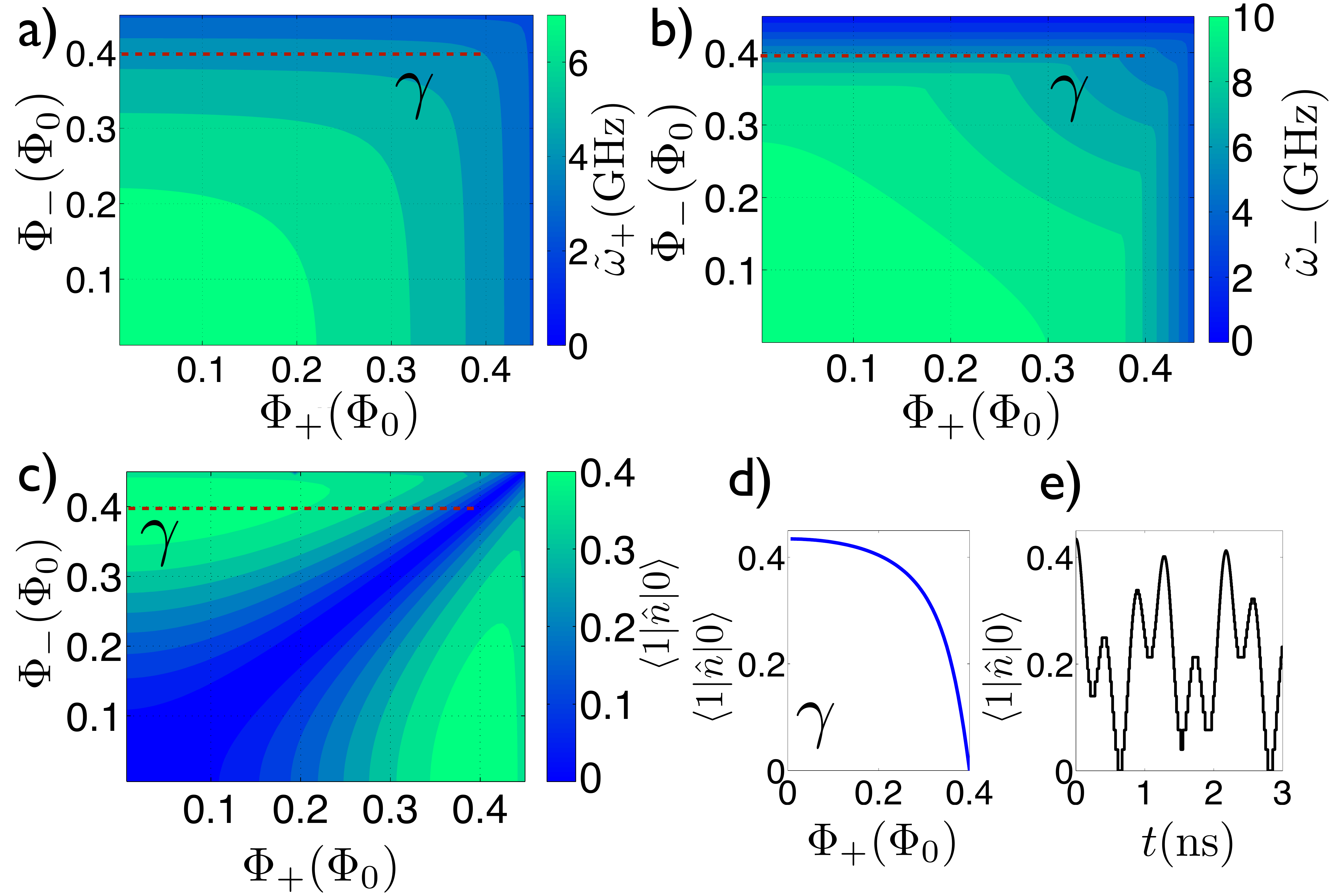}
\caption{(Color online)  a) Transition frequency between the first two levels of the TCQ and b) between the ground state and the third level, as a function of the magnetic fluxes $\Phi_+$ and $\Phi_-$.  c) Matrix element $\langle1|\hat{n}|0\rangle$.  d) Variation of $\langle1|\hat{n}|0\rangle$ along $\gamma$. e) The magnetic flux is varied in time to obtain the time dependence $g_+(t)= g_{+}^s+g_+^d[\cos(\omega_g t)+\cos(\omega'_g t)]$.\label{contour}} 
\end{figure}
The evolution operator associated with the global Hamiltonian in Eq. (\ref{Hresonant}) can be exactly solved, computing a Magnus expansion at second order~\cite{Suppl}. The qubit dynamics gets entangled with the photons in the resonator, and at times $\tau_n=2\pi n /\delta$, with integer $n$, the dynamics is detached from the photons and it follows the Hamiltonian in Eq.~(\ref{HamMS}). The global interaction in Eq. (\ref{HamMS}) is a collective entangling operation between many two level systems. It can be used to obtain many-qubit ~GHZ states at specific times, starting from a configuration in which all the qubits are initialized in the lowest level~\cite{Molmer}. By choosing appropriate initial phases, one can map the dynamics onto $H_{I_{\textrm{eff}}}=-\xi\sum_{i<j} \sigma_i^x\sigma_j^x.$ In general, one can retrieve the dynamics of many-body operators of the form $\sigma^i_1\sigma^j_2\cdots \sigma_N^k$, with $\{ i,j,...k \} \in \{x,y,z\}$~\cite{Mueller11}, up to local qubit rotations. The selectivity upon a generic set of qubits is obtained by setting the coupling between the first two levels to $g_+=0$. The corresponding third level static coupling $g_-=g_-^s$ will not contribute to the dynamics due to the large detuning between the third level and the resonator frequencies. 
\begin{figure}
\includegraphics[scale=0.37]{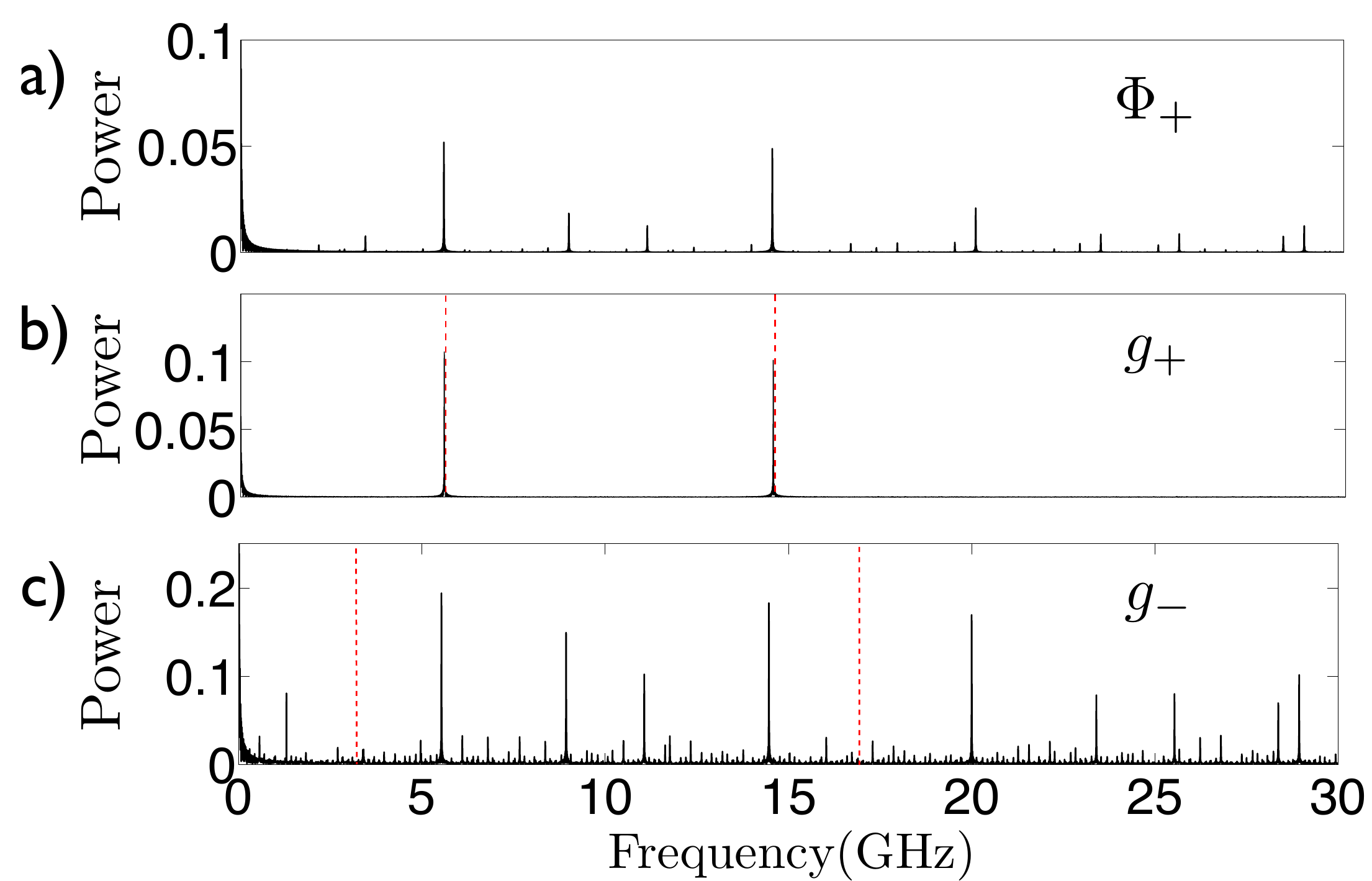}
\caption{(Color online) a) Numerical power spectrum of the magnetic signal $\Phi_+(t)$, used to obtain $g_+(t)$. b) Power spectrum of $g_+(t)$, obtained by plugging the signal $\Phi_+(t)$. The spectrum has two resonances at $\omega_g,\omega'_g$, close to the two sidebands at $\omega_r -\tilde{\omega}_+=5.5$~GHz and $\omega_r+\tilde{\omega}_+=14.5$~GHz (red dotted lines). c) Power spectrum of $g_-(t)$, with the sidebands $\omega_r -\tilde{\omega}_-=3$~GHz and $\omega_r+\tilde{\omega}_-=17$~GHz (red dotted lines). \label{phit}} 
\end{figure}

One can tune in time the coupling, as in Eq. (\ref{tunablecoupling}), by modulating the external magnetic fluxes $\Phi_\pm$. In general, this  will also have an influence on the energy of the first two excited levels. To retain a proper coherent dynamics, one can choose appropriate time-dependent flux drivings such that the coupling has the desired strength, while the qubit transition frequency $\tilde{\omega}_+$ is constant. To give an example, we choose $E_{C_\pm}=500$~MHz,  $E_I=350$~MHz and $E_{J_\pm}=25$~GHz and plot numerically in Fig~\ref{contour}a and~\ref{contour}b, respectively, transition frequencies between the first two levels and the first and the third one, as a function of the flux biases $\{\Phi_+,\Phi_-\}$. For the same parameters, in Fig.~\ref{contour}c, is plotted the matrix element $\langle1|\hat{n}|0\rangle$. Along the curve $\gamma$, approximated by the segment at constant $\Phi_-=0.4\Phi_0$ and $\Phi_+\in [0,0.4] \Phi_0$, the transition frequencies are constant, while $\langle1|\hat{n}|0\rangle$ ranges between a maximum value at $\langle1|\hat{n}|0\rangle^M\simeq0.45$ at $\Phi_+=0$ and a minimum at $\langle1|\hat{n}|0\rangle^m=0$ at $\Phi_+=0.4\Phi_0$, as in Fig.~\ref{contour}d. The coupling range between $\langle1|\hat{n}|0\rangle^M$ and $\langle1|\hat{n}|0\rangle^m$ can be used to encode the time dependent coupling behavior as in Eq.~(\ref{tunablecoupling}). One can design an overall capacitance prefactor $\beta$ such that, e.g., $2eV_{\textrm{rms}}\beta\langle1|\hat{n}|0\rangle^M=g^M_+=80$~MHz. Then one can set $g_+^s\equiv(g_+^M+g_+^m)/2=40$~MHz ($g_+^m=0$) and $g_+^d\equiv(g_+^M-g_+^m)/4=20$~MHz.
By changing $\Phi_+(t)$ along the curve in time, one can encode the proper time-dependence of the coupling.
Notice that the range in which one can drive the magnetic flux is limited by the validity of the negligible charge dispersion regime and by the coupled anharmonic oscillator model, used to describe the TCQ.  In fact, large magnetic fluxes will decrease the effective Josephson energies of the SQUID loops, breaking the regime $E_{J_\pm}\gg E_{C_\pm}$.

Along $\gamma$, one has $\tilde{\omega}_+=4.5$~GHz and $\tilde{\omega}_-=7$~GHz. Furthermore, one can choose $\delta=50$~MHz and consider a resonator frequency of $10$~GHz. The magnetic signal $\Phi_+(t)$ that gives the coupling in Eq.~(\ref{tunablecoupling}) is obtained by inverting the function in Fig~\ref{contour}d, for every time $t$. The coupling, for a sample time interval, is plotted in Fig.~\ref{contour}e. We then decompose the signal $\Phi_+(t)$ in its Fourier components. 
Applying the magnetic signal $\Phi_+(t)$, also the coupling between the first and the third level $g_-(t)$ undergoes fast oscillations. We obtain numerically the time dependence of $g_-(t)$, when the flux $\Phi_+(t)$ is plugged into the system. Considering $g_+^s=40$~MHz, one has a static contribution for $g_-(t)$ of $g_-^s=60$~MHz. The power spectra of $g_+(t)$ and $g_-(t)$ are plotted in Fig~\ref{phit}b and \ref{phit}c. As expected, $g_+(t)$ has only two Fourier components around $\omega_g=14.45$~GHz, $\omega'_g=5.45$~GHz, detuned by $\delta$ from the qubit-resonator sidebands. On the other hand, $g_-(t)$ has no Fourier component close to the resonator-third level sidebands, at $3$~GHz and $17$~GHz. Thus, leakage to the third level of the TCQ will not affect the dynamics. The setup can therefore be regarded as an effective two-level system that undergoes red-detuned and blue-detuned sideband interactions. Furthermore, one can prove that standard Jaynes-Cummings interactions do not affect in a relevant way the dynamics.

\begin{figure}
\includegraphics[scale=0.55]{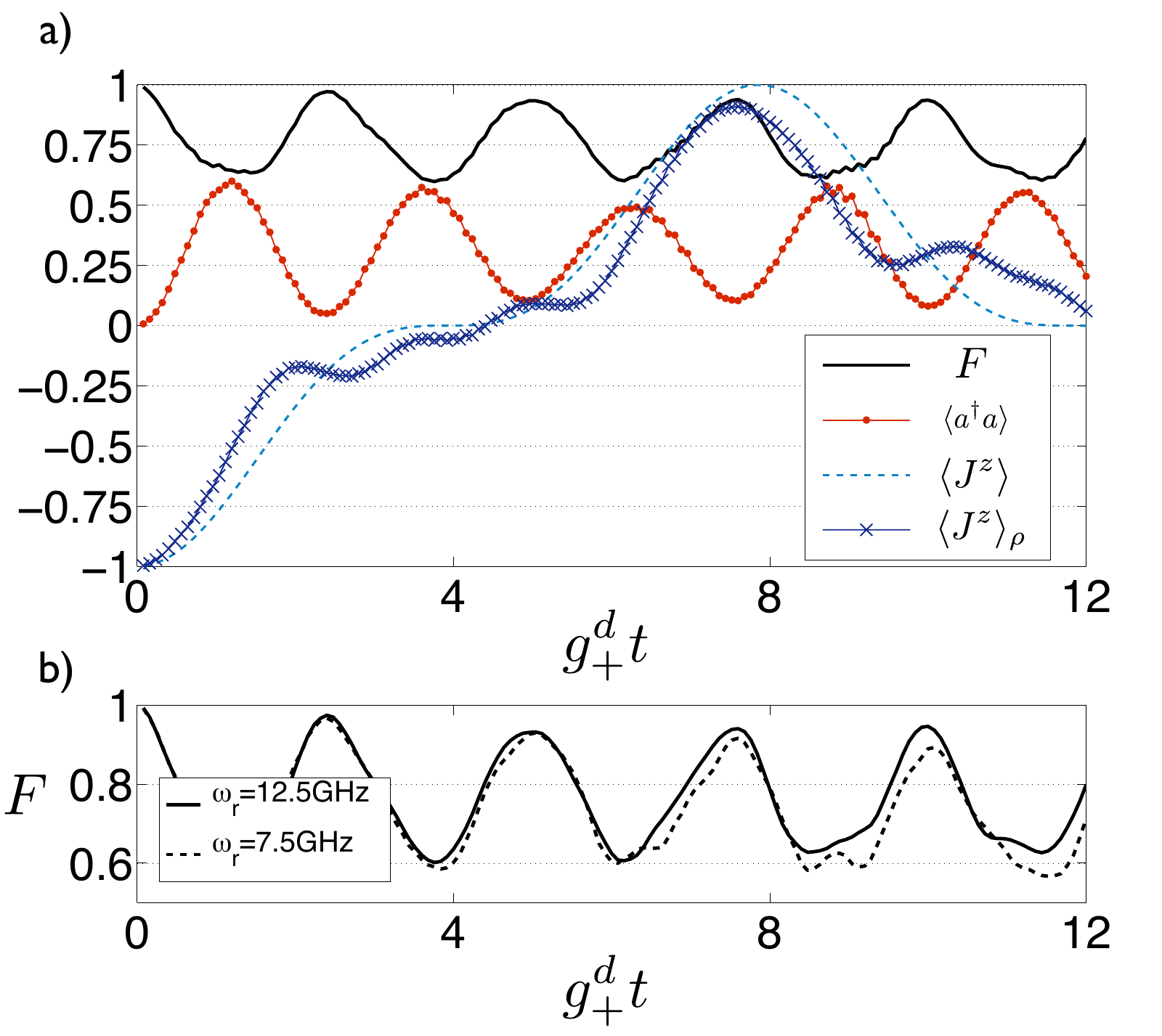}
\caption{(Color online) a) Collective entanglement between four TCQs, intialized in their ground states. The ideal state $|\Psi_I \rangle$ follows the dynamics regulated by the Hamiltonian of Eq.~(\ref{HamMS}), with $\xi=(g_+^d)^2/4\delta$. The fidelity $F=\Tr[{\rho |\Psi_I\rangle\langle\Psi_I |}]$ of the TCQ dynamics is plotted, along with mean number of photons $\langle a^{\dagger}a\rangle$. The ideal mean value of the collective spin oscillation $\langle J_z\rangle$, $J_z=1/4\sum_{i=1}^4\sigma_i^z$, is compared with the TCQ one $\langle J_z\rangle_\rho$. b) Fidelities for different resonator frequencies. The fidelity improves as the qubit-resonator detuning increases. The first two peaks have values $F\approx0.97,0.93$. \label{numsim} }
\end{figure}

Considering that one can maximize the dynamical interaction and choose $g_+^s=2g_+^d$, the condition for neglecting $H_{I_{\textrm{JC}}}$ in Eq.~(\ref{3Ham}), $(g_{+}^s)^2/\Delta_+ \ll (g^d_+)^2/4\delta$, can be formulated in terms of the ratio $\Delta_+/\delta\gg16$. Thus, using higher frequency transitions will improve the fidelity of the gate. To prove this, we perform numerical simulation of the dynamics driven by the interaction Hamiltonian in Eq.~(\ref{3Ham}), in interaction picture with $\tilde{H}_0$. We consider the first three levels for each TCQ. We integrate numerically a Lindblad master equation for the dynamics of four TCQs and resonator, $\dot{\rho}=-i[H_{I_\textrm{eff}},\rho]+\kappa L(a)\rho+\sum_{i=1}^4[\Gamma_\phi L(\sigma^z_i)\rho+\Gamma_- L(\sigma_i^-)\rho]$, adding Lindblad superoperators for the $i$-th qubit $\Gamma_\phi L(\sigma_i^z)\rho$, $\Gamma_- L(\sigma_i^-)\rho$ to take into account dephasing and relaxation rates and $\kappa L(a)\rho$ to take into account resonator losses. Here, $L(A)\rho=(2A\rho A^{\dagger}-A^{\dagger}A\rho-\rho A^{\dagger}A)/2$. We set $\kappa=100$~KHz, $\Gamma_\phi,\Gamma_-=20$~KHz. We use the time-dependent couplings $g_+(t)$, $g_-(t)$ as obtained in Fig.~\ref{phit}. The overall magnitude of the qubit-resonator interaction is set to  $g_+^d=20$~MHz, $g_+^s=40$~MHz, $g_-^s=60$~MHz. We choose $\delta=50$~MHz. The transition frequencies for the first two levels of the TCQ are $\tilde{\omega}_+=4.5$~GHz and $\tilde{\omega}_-=7$~GHz.  The diagonalizing phase $\lambda(t)$ has a fast oscillating contribution. Its effect can be estimated in a  small renormalization of the qubit frequency. In fact, the last term in Eq. (\ref{HTd}) will result, in interaction picture with respect to $\tilde{H}_0$ and neglecting first-order fast-oscillating contribution, into an effective second-order small renormalization of the free energies, leading to $\tilde{\omega}_\pm^R=\tilde{\omega}_\pm+\tilde{\omega}_{\pm \lambda}$, where $\tilde{\omega}_{\pm\lambda}=\lambda_d^2 \omega_\lambda^2 \cdot (\tilde{\omega_+}-\tilde{\omega}_-)/2[(\tilde{\omega}_+ - \tilde{\omega}_-)^2 - \omega_\lambda^2]$ and $\omega_\lambda$ is a frequency of the diagonalizing parameter $\lambda(t)$. The detuning ratio is approximately $\Delta_+/\delta\sim100$. Fig.~\ref{numsim}a shows the fidelity peaks at $\tau_n=2\pi n/\delta$ for the simulated density matrix $\rho$ versus the ideal qubit dynamics, $|\Psi_I\rangle$, that follows the Hamiltonian in Eq.~(\ref{HamMS}), with $\xi=(g^d_+)^2/4\delta$. In Fig.~\ref{numsim}b, the same dynamics is integrated considering two different resonator frequencies. One can notice that, as the qubit-resonator detuning increases, the fidelity peaks get higher as the Jaynes-Cummings part of Eq.~(\ref{3Ham}) is better suppressed.

To perform readout, one can fix $g_+(t)=g_+^s$ and implement standard dispersive measurement with a resonator pull of $\pm(g_+^s)^2/\Delta_+$ depending on the state of the single TCQ~\cite{Blais04,Filipp09}.
 For the practical implementation of this interaction, specific designed flux drivings can take into account inhomogeneous qubit transition frequencies and couplings, by choosing different flux driving trajectories.

In conclusion, we have shown that a setup made out of several superconducting three-island devices, provided with tunable coupling to a coplanar waveguide resonator, may realize collective gates and many-body interactions among superconducting qubits. These interactions can be used to implement topological codes and efficiently simulate fermionic dynamics in circuit QED setups.

We acknowledge useful discussions with Jay Gambetta, Srikanth Srinivasan, and Andreas Wallraff. This work is supported by Basque Government Grant No. IT472-10, Spanish MINECO Grant No. FIS2012-36673-C03-02, Ramón y Cajal Grant No. RYC-2012-11391, UPV/EHU UFI Grant No. 11/55, Swiss National Science Foundation (SNF) Project 150046, CCQED, PROMISCE, and SCALEQIT European projects.

\pagebreak

\begin{widetext}

\section{Supplemental Material for \\ ``Many-Body Interactions with Tunable-Coupling Transmon Qubits''}

In this Supplemental Material, we provide additional details about results of calculations shown in the main text. The effective multiqubit interaction Hamiltonian and the protocol for efficiently obtaining  many-body operators of tunable-coupling transmon qubits are explicitly derived.

\section{Derivation of the effective interaction Hamiltonian}
In this section we show in detail how to derive the effective collective entangling Hamiltonian between $N$ tunable-coupling transmon qubits (TCQs) presented in the manuscript. We start from the interaction between the resonator and several TCQs capacitively coupled to it,
\begin{equation}
\tilde{H}_I=\sum_{j=1}^N\sum_\pm g_\pm(t)(\tilde{b}^{\dagger}_{\pm j} -\tilde{b}_{\pm j})(a^{\dagger}-a).\label{IntSuppl}
\end{equation}
According to what is discussed in the main text, one can design proper magnetic fluxes, threading the SQUIDs in each TCQ, in order to modulate $g_+(t)= g_{+}^s+g_+^d[\cos(\omega_g t)+\cos(\omega'_g t)]$, where one has defined the two detuned sideband frequencies $\omega_g=\omega_r +\tilde{\omega}_+-\delta$ and $\omega'_g=\omega_r -\tilde{\omega}_+ -\delta$. As a consequence, also the transition element to the third level of the devices $g_-(t)$ will undergo fast oscillations. One can numerically obtain its time dependence, and expand the signal in its Fourier components $g_-(t)=\sum_{n} g_n \exp(i\omega_n t)$, with $\omega_n=2\pi n/T$, where $n\in\mathds{Z}$, and $T$ is much larger than the timescale of the dynamics considered. The interaction Hamiltonian in a many qubit setup then becomes 
\begin{eqnarray}
\tilde{H}_I=&&\sum_{j=1}^N\left[g_{+}^s+g_+^d\left(\cos(\omega_g t)+\cos(\omega'_g t)\right)\right](\tilde{b}^{\dagger}_{+j}-\tilde{b}_{+j})(a-a^{\dagger})\nonumber\\
&&+\sum_{j=1}^N\left[\sum_{n} g_n \exp(i\omega_n t)\right](\tilde{b}^{\dagger}_{-j} -\tilde{b}_{-j})(a-a^{\dagger}).\label{Hspectrum}
\end{eqnarray}
One can identify three contributions to the dynamics, $\tilde{H}_I=  \tilde{H}_{I_{\textrm{JC}}} + \tilde{H}_{I_+} + \tilde{H}_{I_-}$. There are two terms representing standard Jaynes-Cummings interactions, due to the static contributions of the couplings,
 \begin{equation}
\tilde{H}_{I_\textrm{JC}}=-\sum_{j=1}^N\sum_\pm g_{\pm}^s(\tilde{b}^{\dagger}_{\pm j}a+\tilde{b}_{\pm j}a^{\dagger}),
\end{equation}
where we have defined $ g_-^s \equiv g_0$. The other contributions to the dynamics are given by the time-dependent part of the interaction. Namely,
\begin{eqnarray}
\tilde{H}_{I_+}&&=\sum_{j=1}^Ng_+^d\left[\cos(\omega_g t)+\cos(\omega'_g t)\right](\tilde{b}^{\dagger}_{+j}-\tilde{b}_{+j})(a-a^{\dagger}),\nonumber\\
\tilde{H}_{I_-}&&=\sum_{j=1}^N\left[\sideset{}{'}\sum_{n}   g_n \exp(i\omega_n t)\right](\tilde{b}^{\dagger}_{-j} -\tilde{b}_{-j})(a-a^{\dagger}),\label{Hspectrum2}
\end{eqnarray}
where the prime symbol excludes the zeroth addend from the series. One can define an interaction picture with respect to $\tilde{H}_0$, see Eq.~(1) in the main text, and neglect the $\tilde{H}_{I_-}$ contribution, if there is no large component $g_n$ of the Fourier decomposition, whose frequency $\omega_n$ is close to the resonator-third level sidebands. This is shown to be the case in Fig. 3 in the main text. Due to sufficient level anharmonicity, that  one can assume being preserved during the dynamics, only the two lowest levels for each anharmonic oscillator are populated. One is thus allowed to consider a two-level Pauli algebra to model qubit excitations, $\tilde{b}_{\pm j}\equiv\sigma_{\pm j}^-$(equivalently $\tilde{b}_{\pm j}^{\dagger}\equiv\sigma_{\pm j}^+$), and the interaction Hamiltonian, in the rotated frame, becomes
\begin{equation}
\tilde{H}_I\approx\sum_{j=1}^N\left[g_{+}^s+g_+^d\left(\cos(\omega_g t)+\cos(\omega'_g t)\right)\right](\sigma_{+j}^+e^{i\tilde{\omega}_+t}-\sigma_{+ j}^-e^{-i\tilde{\omega}_+t})(ae^{-i\omega_rt}-a^{\dagger}e^{i\omega_rt}).
\end{equation}
Under the condition $|(g_{+}^s)^2/\Delta_+ |\ll |(g^d_+)^2/4\delta|$, the biggest contribution to the dynamics come from the terms rotating at the smallest frequency $\delta$,
\begin{equation}
\tilde{H}_I=\sum_{j=1}^N\frac{g_+^d}{2}\left\{(\sigma^+_{+j}-\sigma^-_{+j})(a^{\dagger}e^{i\delta t}-ae^{-i\delta t})\right\} =-i\frac{g_+^d}{2}S^y(a^{\dagger}e^{i\delta t}-ae^{-i\delta t}),\label{MSCQED}
\end{equation}
where $S^y=\sum_{j=1}^N\sigma_{+j}^y$. The evolution operator associated with Hamiltonian in Eq.~(\ref{MSCQED}) can be computed exactly at second order in $g_+^d/2$, obtaining 
\begin{equation}
\tilde{U}_I(t)=\exp\left\{\frac{g_+^d S^y}{2\delta}\left[(e^{i\delta t}-1)a^{\dagger}-\textrm{H.c.}\right]\right\}\exp\left\{i\left(\frac{g_+^d}{2\delta}(S^y)\right)^2\left[\sin(\delta t)-\delta t\right]\right\}.
\end{equation}
At times $\tau=2\pi n/\delta$, with integer $n$, the above evolution operator can be associated with the effective unitary  $\tilde{U}_I(t)=\exp\left[{i\sum_{ij}(g_+^d)^2/4\delta\sigma_{+i}^{y}\sigma_{+j}^{y}}\right]$. By choosing appropriate initial phases in Eq.~(\ref{IntSuppl}), one can obtain the generic effective interaction (here $\alpha=\{x,y\}$)
\begin{equation}
H_{I_{\textrm{eff}}}=-\sum_{ij}\frac{(g_+^d)^2}{4\delta}\sigma_{+i}^{\alpha}\sigma_{+j}^{\alpha}.\label{HMSSuppl}
\end{equation}

\section{Effective Many-Body Operators}
In this section we show explicitly, starting from the effective Hamiltonian in Eq.~(\ref{HMSSuppl}), how to obtain an effective many-body interaction of $N$ qubits, along the lines of Refs.~\cite{Mueller11S, Barreiro11S}.  We consider a combination of direct and inverse collective gates and a local rotation on one of the qubits (e.g. the first one). In other words, we consider the gate sequence $U_\textrm{S}(t)=\exp(- i H_{I_\textrm{eff}} \tau)\exp(igt\sigma^z_1)\exp(i H_{I_\textrm{eff}} \tau)$, where $\tau=\phi2\delta/(g_+^d)^2$, that explicitly reads
\begin{equation}
U_\textrm{S}(t,\phi)=e^{i\phi/2\sum_{j=2}\sigma^\alpha_i\sigma^\alpha_j}e^{igt\sigma^z_1}e^{-i\phi/2\sum_{j=2}\sigma^\alpha_i\sigma^\alpha_j}.
\end{equation}
One can expand the local rotation and write the equivalent expression
\begin{equation}
U_\textrm{S}(t,\phi)=e^{i\phi/2\sigma^\alpha_1\sum_{j=2}\sigma^\alpha_j}\left(\cos(gt)+i\sin(gt)\sigma^z_1\right)e^{-i\phi/2\sigma^\alpha_1\sum_{j=2}\sigma^\alpha_j}.
\end{equation}
Taking into account that $\sigma^z_1e^{i\phi/2\sigma^\alpha_1\sum_{j=2}\sigma^\alpha_j}=e^{-i\phi/2\sigma^\alpha_1\sum_{j=2}\sigma^\alpha_j}\sigma^z_1$,
one has that
\begin{equation}
U_\textrm{S}(t,\phi)=\cos(gt)+i\sin(gt)\sigma^z_1e^{-i\phi\sigma^\alpha_1\sum_{j=2}\sigma^\alpha_j}\label{eq:2}.
\end{equation}
Considering that $\left(\sigma^z_1e^{-i\phi/2\sigma^\alpha_1\sum_{j=2}\sigma^\alpha_j}\right)^{n}=\{1,\sigma^z_1e^{-i\phi/2\sigma^\alpha_1\sum_{j=2}\sigma^\alpha_j}\}$ for $n=\{\textrm{even,odd}\}$, Eq. (\ref{eq:2}) can be rewritten,
\begin{equation}
U_S(t,\phi)=\exp\left(igt\sigma^z_1\prod_{j=2}\left(\cos(\phi)-i\sigma^\alpha_1\sigma^\alpha_j\sin(\phi)\right)\right).
\end{equation}
Choosing $\phi=\pi/2$, one has
\begin{equation}
U_\textrm{S}(t,\pi/2)=\exp\left(igt\sigma^z_1\prod_{j=2}\left(-i\sigma^\alpha_1\sigma^\alpha_j\right)\right).
\end{equation}

\begin{figure}
\includegraphics[scale=0.5]{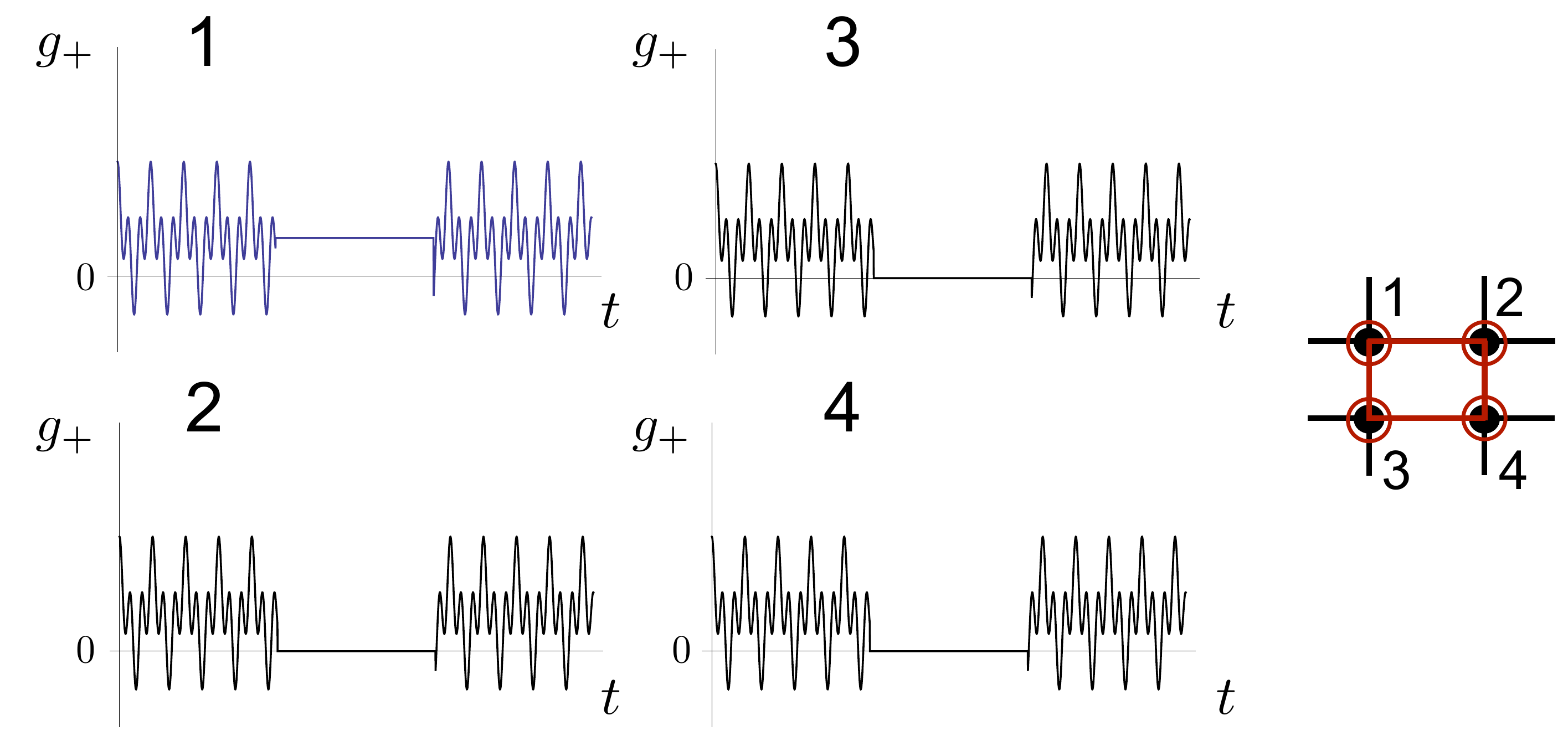}
\caption{(Color online) a) Scheme of the generation of many-particle operator among four TCQs. The coupling of the four qubits to the resonator is shown as a function of time. Collective gates as in Eq.~(\ref{HMSSuppl}) are performed in the initial and final time regions, while a standard phase gate is performed upon the first qubit between the two collective operations. The effective interaction can be mapped on an arbitrary stabilizer operator on a spin lattice with generic topology, due to the non-local nature of the quantum bus. With an additional ancillary qubit, the system state can be mapped on the ground states of topological codes, via stabilizer pumping. \label{4QubitGateScheme}} 
\end{figure}

The resulting gate, as a function of the total number of qubits $N$, reads
\begin{align}
&\exp\left(-igt\sigma^z_1\sigma_{2}^{\alpha}\cdots\sigma_{N}^{\alpha}\right), N=4n-1,\nonumber\\
&\exp\left(igt\sigma^z_1\sigma_{1}^{\alpha}\cdots\sigma_{N}^{\alpha}\right), N=4n+1,\nonumber\\
&\exp\left(igt\sigma^\beta_1\sigma_{2}^{\alpha}\cdots\sigma_{N}^{\alpha}\right), N=4n,\nonumber\\
&\exp\left(-igt\sigma^\beta_1\sigma_{2}^{\alpha}\cdots\sigma_{N}^{\alpha}\right), N=4n-2,\label{stabilizer}
\end{align}
where $\sigma_1^\beta=-\sigma_1^y(\sigma_1^x)$ for $\alpha=x(y),N=4n$, and $\sigma_1^\beta=\sigma_1^y(-\sigma_1^x)$ for $\alpha=x(y),N=4n-2$. All these interactions are equivalent to an arbitrary stabilizer many-body operator, up to local rotations. Summarizing, the physical realization of the multiqubit interaction can be schematized as in Fig.~\ref{4QubitGateScheme}. Magnetic fluxes drive the collective gates at the beginning and the end of the protocol, while in the central time interval the coupling with the resonator of all the qubits is turned off, except for the TCQ that undergoes a standard phase shift gate (qubit 1 in the figure). By adding an auxiliary ancilla qubit one can guide the ground state of the system to the one of topological states~\cite{Kitaev03}, via the stabilizer pumping protocol described in~\cite{Barreiro11S}.
Sequences of collective operators as in Eq.~(\ref{stabilizer}) can be used to simulate correlated fermionic Hamiltonians in spin systems, with a constant overhead of the quantum resources, according to the protocols presented in~\cite{Casanova12,Mezzacapo12}.

\end{widetext}


\begin{thebibliography}{99}

\bibitem{Devoret13} M. H. Devoret and R. J. Schoelkopf, Science {\bf 339}, 1169 (2013).

\bibitem{Wallraff04}A. Wallraff, D. I. Schuster, A. Blais, L. Frunzio, R.-S. Huang, J. Majer, S. Kumar, S. M. Girvin, and R. J. Schoelkopf, Nature {\bf 431}, 162 (2004).

\bibitem{Steffen06} M. Steffen, M. Ansmann, R. C. Bialczak, N. Katz, E. Lucero, R. McDermott, M. Neeley, E. M. Weig, A. N. Cleland, and J. M. Martinis, Science {\bf313}, 1423 (2006).

\bibitem{Fedorov12}A. Fedorov, L. Steffen, M. Baur, M. P. da Silva, and A. Wallraff, Nature {\bf481}, 170 (2012).

\bibitem{Steffen13}L. Steffen, Y. Salathe, M. Oppliger, P. Kurpiers, M. Baur, C. Lang, C. Eichler, G. Puebla-Hellmann, A. Fedorov, and A. Wallraff, Nature {\bf 500}, 319 (2013).

\bibitem{Niemczyk10} T. Niemczyk, F. Deppe, H. Huebl, E. P. Menzel, F. Hocke, M. J. Schwarz, J. J. Garc\'ia-Ripoll, D. Zueco, T. H\"ummer, E. Solano, A. Marx	, and R. Gross, Nat. Phys. {\bf 6}, 772 (2010).

\bibitem{Peropadre10} B. Peropadre, P. Forn-D\'iaz, E. Solano, and J. J. Garc\'ia-Ripoll, Phys. Rev. Lett. {\bf 105}, 023601 (2010).

\bibitem{Koch07} J. Koch, Terri M. Yu, J. Gambetta, A. A. Houck, D. I. Schuster, J. Majer, A Blais, M. H. Devoret, S. M. Girvin and R. J. Schoelkopf, Phys. Rev. A, {\bf 76}, 042319 (2007)

\bibitem{Reed12} M. D. Reed, L. DiCarlo, S. E. Nigg, L. Sun, L. Frunzio, S. M. Girvin and R. J. Schoelkopf, Nature {\bf 482}, 382 (2012).

\bibitem{Abdumalikov13} A. A. Abdumalikov, J. M. Fink, K. Juliusson, M. Pechal, S. Berger, A. Wallraff and S. Filipp, Nature {\bf 496}, 482 (2013).

\bibitem{Heras14} U. Las Heras, A. Mezzacapo, L. Lamata, S. Filipp, A. Wallraff, and E. Solano, Phys. Rev. Lett. {\bf 112}, 200501 (2014).

\bibitem{Mezzacapo14} A. Mezzacapo, U. Las Heras, J. S. Pedernales, L. DiCarlo, E. Solano, and L. Lamata, e-print arXiv:1405.5814.

\bibitem{Bishop09} L. S. Bishop, L. Tornberg, D. Price, E. Ginossar, A. Nunnenkamp, A. A. Houck, J. M. Gambetta, J. Koch, G. Johansson, S. M. Girvin and R. J. Schoelkopf, New J. Phys. {\bf11}, 073040 (2009).

\bibitem{Neeley10} M. Neeley, R. C. Bialczak, M. Lenander, E. Lucero, M. Mariantoni, A. D. OÕConnell, D. Sank, H. Wang, M. Weides, J. Wenner,	 Y. Yin, T. Yamamoto, A. N. Cleland and J. M. Martinis, Nature {\bf467}, 570 (2010). 

\bibitem{DiCarlo10}L. DiCarlo, M. D. Reed, L. Sun, B. R. Johnson, J. M. Chow, J. M. Gambetta, L. Frunzio, S. M. Girvin, M. H. Devoret and R. J. Schoelkopf, Nature {\bf467}, 574 (2010).

\bibitem{Molmer} K. M\o lmer and A. S\o rensen, Phys. Rev. Lett. {\bf 82}, 1835-1838 (1999); A. S\o rensen and K. M\o lmer, Phys. Rev. A, {\bf 62}, 022311 (2000).

\bibitem{Monz11} T. Monz, P. Schindler, J. T. Barreiro, M. Chwalla, D. Nigg, W. A. Coish, M. Harlander, W. H\"ansel, M. Hennrich and R. Blatt, Phys. Rev. Lett. {\bf 106}, 130506 (2011).

\bibitem{Liu07} Y. X. Liu, L. F. Wei, J. R. Johansson, J. S. Tsai and F. Nori, Phys. Rev. B {\bf 76}, 144518 (2007).

\bibitem{Wallraff07} A. Wallraff, D. I. Schuster, A. Blais, J. M. Gambetta, J. Schreier, L. Frunzio, M. H. Devoret, S. M. Girvin and R. J. Schoelkopf, Phys. Rev. Lett. {\bf 99}, 050501 (2007).

\bibitem{Leek08} P. J. Leek, S. Filipp, P. Maurer, M. Baur, R. Bianchetti, J. M. Fink, M. G\"ošppl, L. Steffen, A. Wallraff, Phys. Rev. B {\bf 79}, 180511(R) (2009).

\bibitem{Strand13} J. D. Strand, Matthew Ware, FŽlix Beaudoin, T. A. Ohki, B. R. Johnson, Alexandre Blais, B. L. T. Plourde, Phys. Rev. B {\bf 87}, 220505(R) (2013).

\bibitem{Mueller11}M. M\"uller, K. Hammerer, Y. L. Zhou, C. F. Roos and P. Zoller, New J. Phys. {\bf 13}, 085007 (2011); J.T. Barreiro, M. M\"uller, P. Schindler, D. Nigg, T. Monz, M. Chwalla, M. Hennrich, C. F. Roos, P. Zoller and R. Blatt, Nature {\bf 470}, 486 (2011). 

\bibitem{Nigg12}S. E. Nigg and S. M. Girvin, Phys. Rev. Lett. {\bf 110}, 243604 (2013).

\bibitem{Kitaev03} A. Kitaev, Ann. Phys., {\bf 303}, 2 (2003).

\bibitem{Casanova12}J. Casanova, A. Mezzacapo, L. Lamata and E. Solano, Phys. Rev. Lett. {\bf 108}, 190502 (2012).

\bibitem{Mezzacapo12}A. Mezzacapo, J. Casanova, L. Lamata and E. Solano, Phys. Rev. Lett. {\bf 109}, 200501 (2012).

\bibitem{Gambetta11} J. M. Gambetta, A. A. Houck and A. Blais, Phys. Rev. Lett. {\bf 106}, 030502 (2011).

\bibitem{Srinivasan12} S. J. Srinivasan, A. J. Hoffman, J. M. Gambetta and A. A. Houck, Phys. Rev. Lett. {\bf 106}, 083601 (2011).

\bibitem{Blais04}A. Blais, R.-S. Huang, A. Wallraff, S. M. Girvin, and R. J. Schoelkopf, Phys. Rev. A, {\bf 69}, 062320 (2004).

\bibitem{Filipp09}S. Filipp, P. Maurer, P. J. Leek, M. Baur, R. Bianchetti, J. M. Fink, M. G\"oppl, L. Steffen, J. M. Gambetta, A. Blais, and A. Wallraff,
Phys. Rev. Lett. {\bf 102}, 200402 (2009).

\bibitem{Suppl} See provided supplemental material for further analysis and discussions of the results in the main text.

\end{thebibliography}

\begin{thebibliography}{99}

\bibitem{Mueller11S} M. M\"uller, K. Hammerer, Y. L. Zhou, C. F. Roos and P. Zoller, New J. Phys. {\bf 13}, 085007 (2011); 

\bibitem{Barreiro11S}J.T. Barreiro, M. M\"uller, P. Schindler, D. Nigg, T. Monz, M. Chwalla, M. Hennrich, C. F. Roos, P. Zoller and R. Blatt, Nature {\bf 470}, 486 (2011). 

\bibitem{Kitaev03} A. Yu. Kitaev, Ann. Phys. {\bf 303}, 2 (2003).

\bibitem{Casanova12}J. Casanova, A. Mezzacapo, L. Lamata and E. Solano, Phys. Rev. Lett. {\bf 108}, 190502 (2012).

\bibitem{Mezzacapo12}A. Mezzacapo, J. Casanova, L. Lamata and E. Solano, Phys. Rev. Lett. {\bf 109}, 200501 (2012).

\end{thebibliography}
\end{document}